\documentclass[a4paper,10pt,twoside,utf8]{cpc-hepnp}
\usepackage{multicol}
\usepackage{graphicx}
\usepackage{amssymb,bm,mathrsfs,bbm,amscd}
\usepackage[tbtags]{amsmath}
\usepackage{lastpage}
\usepackage{CJK}
\usepackage{epstopdf}
\usepackage{epsfig}
\usepackage{color}
\usepackage{ifpdf}
\usepackage{cuted}
\usepackage{caption}
\usepackage{url}
\usepackage{tabularx}

\usepackage{adjustbox}
\usepackage{multirow}

\usepackage[colorlinks,linkcolor=blue,citecolor=blue]{hyperref}
\usepackage{booktabs}    
\usepackage[normalem]{ulem}

\begin{document}

\fancyhead[c]{\small Chinese Physics C~~~Vol. {49}, No. 6 (2025) 064104} \fancyfoot[C]{\small 064104-\thepage}
\title{Configuration-interaction relativistic Hartree-Fock model}
\author{
      Jia Liu $^{1,2)}$
\quad Yi Fei Niu $^{1,2,3,4)}$
\quad Wen Hui Long $^{1,2,4)}$\thanks{Email: longwh@lzu.edu.cn}
}
\maketitle
\footnotetext{The authors thank Dr. P.W. Zhao and Dr. F.Q. Chen for their fruitful discussions. This work is partly supported by the Strategic Priority Research Program of the Chinese Academy of Sciences under Grant No. XDB34000000, the Fundamental Research Funds for the Central Universities (Grant No. lzujbky-2023-stlt01), the National Key Research and Development (R$\&$D) Program under Grant No. 2021YFA1601500, the National Natural Science Foundation of China under Grant Nos. 1240050235 and 12275111, and the Supercomputing Center of Lanzhou University. }
\address{%
 $^1$Frontier Science Center for Rare isotope, Lanzhou University, Lanzhou 730000, China\\
 $^2$School of Nuclear Science and Technology, Lanzhou University, Lanzhou 730000, China\\
 $^3$Department of Nuclear Physics, China Institute of Atomic Energy, Beijing, 102413, China\\
 $^4$Joint Department for Nuclear Physics, Lanzhou University and Institute of Modern Physics, CAS, Lanzhou 730000, China
}

\begin{abstract}
 The configuration interaction relativistic Hartree-Fock (CI-RHF) model is developed in this work. Compared to the conventional configuration interaction shell model (CISM), the CI-RHF model can be applied to study the structural properties of a wide range of nuclei without readjusting any parameters, as the effective Hamiltonian for different model space can be deduced consistently from a universal density-dependent Lagrangian based on the Hartree-Fock single-particle basis. The convergence of intermediate-state excitations has been examined in evaluating the effective interactions, and the core-polarization effects are illustrated, by using $^{18}$O as an example. Employing the CI-RHF model, both the bulk properties and low-lying spectra of even-even nuclei $^{18\sim 28}$Ne have been well reproduced with the model space restricted to the $sd$ shell. Studies of the isotopic evolution concerning charge radii and low-lying spectra highlight the shell closure at $N=14$ for neon isotopes. Furthermore, the cross-shell calculations extending from the $sd$ to $pf$ shell successfully reproduced the low-lying spectra of $^{30}$Ne and $^{32}$Ne. Notably, remarkably low excitation energies $E(2^{+}_{1})$ of $^{30}$Ne suggest the disappearance of the conventional magicity $N=20$.
\end{abstract}

{\small\ {\bfseries DOI}: 10.1088/1674-1137/adbdba \hspace{2em}{\bfseries CSTR}: 32044.14.ChinesePhysicsC.49064104}

\footnotetext[0]{\hspace*{-3mm}\raisebox{0.3ex}{$\scriptstyle\copyright$}2025Chinese Physical Society and the Institute of High Energy Physics of the Chinese Academy of Sciences and the Institute of Modern Physics of the Chinese Academy of Sciences and IOP Publishing Ltd. All rights, including for text and data mining, AI training, and similar technologies, are reserved.}%

\begin{multicols}{2}

\section{INTRODUCTION}\label{INTRODUCTION}
During the past decades, benefiting from the worldwide development of rare isotope beam facilities, a multitude of novel nuclear phenomena have been discovered, such as halo structures \cite{Tanihata1985PRL55, Kobayashi2014PRL112}, new magic numbers \cite{Ozawa2000PRL84, Steppenbeck2013Nature502, Wienholtz2013Nature498}, and the so-called ``island of inversion'' \cite{Warburton1990PRC41, Nowacki2021PPNP120}. These discoveries have significantly extended the interests of nuclear physicists from stable nuclides to unstable ones. Meanwhile, these also present substantial challenges to our understanding of nuclear force and the methodology employed to address complex many-body nuclear system.

Configuration interaction shell model (CISM) has been one of the most powerful methods to understand the fundamental properties of nuclear structure, especially the sophisticated low-lying spectra \cite{Caurier2005RMP77}. Considering that the dimensionality of the configuration space grows exponentially with the increase of mass number, the CISM calculations are typically constrained to a truncated Hilbert space, namely the model space, where the core is frozen and only the degrees of freedom associated with valence nucleons are considered. As a result, an effective Hamiltonian should be introduced to take into account the configurations excluded from the model space. Practically, the success of shell model essentially depends on the choice of the configuration interactions \cite{Brown1988AP182,Dean2004PPNP53,Stroberg2019ARNS69}.

Phenomenologically, the effective Hamiltonian, tailored to a specific model space, can be obtained by fitting the selected experimental data, e.g., the USD family of Hamiltonian for the $sd$ shell \cite{Wildenthal1984PPNP11, Brown2006PRC74}. The fully microscopic derivation of the effective Hamiltonian based on the realistic forces remains a long-standing challenge for shell model calculations \cite{Hjorth1995PR261, Coraggio2009PPNP62, Hergert2016PR621, Stroberg2019ARNS69}. Following time-dependent or time-independent perturbation theory \cite{Goldstone1957, BH1967RMP39}, the $\widehat{Q}$-box resummation method, also known as the folded diagram approach \cite{Kuo1974AR24, Kuo1990FDT}, has been widely used to derive the shell model Hamiltonian. It begins with the $G$-matrix or low-momentum nucleon-nucleon (NN) potential $V_{\text{low-}k}$ to treat the strong short-range correlations induced by the realistic NN force \cite{Brueckner1955PR100, Bogner2003PR386}, then the $\widehat{Q}$-box corrections and the folded terms are evaluated to account for the core polarization and other important correlations \cite{Brown1967NPA92, Bogner2002PRC65}.

Shell model calculations with such effective Hamiltonians present a satisfactory description of the binding energies and low-lying excitations of the nuclei with a few valence particles \cite{Kuo1966NP85, Kuo1968NPA114, Herling1972NPA181}, while the agreement with experimental data deteriorates as more valence nucleons are added \cite{McGrory1970PRC2, Cole1975JPG1, Coraggio2009PPNP62}. In fact, the main deficiency of such interactions lies in the monopole part \cite{Zuker2003PRL90}, which is associated with the mean field of the specific nucleus. To improve the accuracy of theoretical descriptions, an empirical mass dependence of two-body interaction matrix elements (TBMEs) is typically adopted both for microscopic and phenomenological shell model calculations \cite{Wildenthal1984PPNP11, Brown1988AP182,Honma2004PRC69, Tsunoda2017PRC95}. Notably, the effects that account for the mass dependence of the TBMEs are nontrivial. On the one hand, the nuclear mean field changes consistently with the addition of valence nucleons, which indicates that the single-particle wave functions are mass dependent \cite{Brown1988AP182, Wildenthal1984PPNP11}. By comparing the $G$-matrix obtained for different nuclei, Hosaka {\itshape et al.} pointed out that the $G$-matrix elements also exhibit a simple mass dependence \cite{Hosaka1985NPA444}. Besides that, an important missing part of such monopole interactions is thought to be contributed by the three-body forces \cite{Zuker2003PRL90, Otsuka2010PRL105, Stroberg2017PRL118}. Considering the difficulty in handling three-body Hamiltonians for shell model diagonalization, the dominant contributions of the three-body forces are typically incorporated into a density-dependent two-body term \cite{Polls1983NPA401, Holt2013EPJA49, Tsunoda2017PRC95}.

In contrast to the CISM, with a given inter-particle interaction, the nuclear energy density functional (EDF) or the Hartree-Fock (HF) model can be applied to the study of nuclei across nearly the entire nuclide chart \cite{Vautherin1972PRC5, Gogny1980PRC21, Reinhard1989RPP52}. In the EDF or HF framework, the ground state properties of the specific nucleus are determined by a self-consistent mean field or HF potential, which is associated with the one-body density and can be derived through a variational procedure. Correspondingly, the essential physics encapsulated within the mean field or HF potential is manifested in the single-particle structure.

On the other hand, the empirical inter-particle interactions used in the EDF or HF model are obtained by fitting the properties of nuclei and nuclear matter at the saturation point \cite{Typel1999NPA656, Stone2007PPNP58, Robledo2018JPG46}. These interactions can be considered as a kind of phenomenological $G$-matrix \cite{Vautherin1972PRC5, Negele1987PRC5}, and their monopole and multipole parts are typically density-dependent to account for the nuclear in-medium effects \cite{Skyrme1958NP9, Brockmann1992PRL68}. Notably, recent intensive works related to unstable nuclei have highlighted the importance of non-central parts of nuclear force, namely the spin-orbit and tensor forces, in the shell evolution \cite{Otsuka2005PRL95, Lesinski2007PRC76, Wang2018PRC98}, the isotopic shifts of charge radii \cite{Sharma1995PRL74, Nakada2015PRC91, Nakada2019PRC100}, and the spin-isospin excitations \cite{Bai2010PRL105, Wang2020PRC101, Sagawa2014PPNP76}, etc. Compared to the non-relativistic EDF, the spin-orbit force is naturally included in the relativistic scheme. Furthermore, the tensor force can be explicitly obtained via the exchange (Fock) terms in the relativistic Hartree-Fock (RHF) model \cite{Long2006PLB640, Long2007PRC76, Long2010PRC81, Long2008EPL82, Wang2013PRC87, Jiang2015PRC91, Wang2018PRC98, Geng2020PRC101}. 

With several fitted parameters, the empirical EDFs or HF models and their extensions have achieved satisfactory descriptions of various nuclear phenomena from light to superheavy nuclei \cite{Meng2006PPNP57, Delaroche2010PRC81, Kortelainen2010PRC82, Niksic2011PPNP66, Erler2012Nature486, Zhang2022ADNDT144}. Naturally, one may expect that the density-dependent empirical interactions and the self-consistent single-particle basis given by the EDF or HF model could provide a reliable and general input for the shell model \cite{Waroquier1983NPA404, Sagawa1985PLB159,Gomez1993NPA551, Bender1996NPA596, Brown1998PRC58, Rodriguez2008PRC77, Brown2011PLB695, Zhao2016PRC94, Jiang2018PRC98}, where the effective Hamiltonians suffer from the issues of locality and mass dependence. In fact, Brown and the collaborators suggested that the mass-dependent monopole terms in the CISM calculations can be constrained by the EDF results, which contain the three-body, density-dependent and rearrangement contributions \cite{Brown2011PLB695}. At the earlier time, Waroquier and the collaborators proposed a shell model approach based on the self-consistent Hartree-Fock single-particle basis with a universal Skyrme-type force \cite{Waroquier1983NPA404}. The calculated two-particle (hole) excited spectra of doubly magic nuclei are in agreement with experimental results, with some adjusted or empirical single-particle energies. Their studies have shown that the effective TBMEs, derived from the Skyrme force SKE, and the calculated low-lying spectra strongly depend on the strength of the density-dependent zero-range force.

In this paper, we aim to integrate the strengths of the CISM and RHF theory to achieve a more robust method, namely the configuration-interaction relativistic Hartree-Fock (CI-RHF) model. On the one hand, we describe the nuclear system with an empirical but general Lagrangian, in which the spin-orbit and tensor forces are naturally included, and the nuclear in-medium effects are evaluated phenomenologically via the density dependencies of the meson-nucleon coupling strengths \cite{Long2006PLB640, Long2007PRC76, Long2010PRC81, Geng2020PRC101, Geng2022PRC105}. On the other hand, the RHF calculations provide an optimized single-particle basis to construct the model space, and in such a way, the energy of the frozen core can be calculated consistently. Following the spirit of the CISM, we derive the effective Hamiltonian self-consistently using the folded-diagram approach, and then diagonalize it in the model space to obtain the ground state and low-lying excitation properties. It is expected that the CI-RHF model can be applied to study the properties of a wide range of nuclei, without introducing additional parameters besides those well-defined in the phenomenological Lagrangian.

The paper is organized as follows. In Sec. {\ref{FORMALISM}}, we present the general formalism of the CI-RHF model, and the calculation details of the effective Hamiltonian are provided in Appendix {\ref{APPENDIX}}. In Sec. {\ref{RESULTS}}, we first examine the convergence of intermediate-state excitations in calculating the effective TBMEs and illustrate the core-polarization effects. Subsequently, taking $^{16}$O, $^{40}$Ca and $^{48}$Ca as examples, we investigate the beyond-mean-field corrections to the doubly magic nuclei. Furthermore, we obtain the properties of both ground states and low-lying excited states for neon isotopes with different model space, demonstrating the self-consistency of the CI-RHF model. Finally, a brief summary is given in Sec. \ref{CONCLUSION}.

\section{GENERAL FORMALISM}\label{FORMALISM}
Based on the meson propagated diagram of nuclear force \cite{Yukawa1935PPMSJ17}, the Lagrangian of nuclear systems can be established considering the degrees of freedom associated with the nucleon field ($\psi$), the isoscalar scalar $\sigma$- and vector $\omega$-meson fields, the isovector vector $\rho$- and pseudo-scalar $\pi$-meson fields, and the photon field ($A$) \cite{Bouyssy1987PRC36, Long2006PLB640, Long2010PRC81},
\begin{equation}\label{Lagrangian}
	\mathscr{L} = \mathscr{L}_{\psi} + \mathscr{L}_{\sigma} + \mathscr{L}_{\omega} + \mathscr{L}_{\rho} + \mathscr{L}_{\pi} + \mathscr{L}_{A} + \mathscr{L}_{I},
\end{equation}
where $\mathscr{L}_{I}$ contains the interactions between the nucleon and meson (photon) fields, and $\mathscr L_\phi$ ($\phi = \psi, \sigma, \omega, \rho, \pi$, and $A$) represents the free parts of nucleon and meson (photon) fields. Starting from the above Lagrangian, the Hamiltonian of nuclear systems can be derived from the Legendre transformation, and further the energy functional can be obtained following the standard procedure as detailed in Refs. \cite{Bouyssy1987PRC36, Long2006PLB640, Geng2020PRC101, Geng2022PRC105}.

Restricted to the Hartree-Fock level, nucleons are treated as independent particles moving in an averaged field. Considering that collective correlations are significantly hindered by the shell structure, such an independent particle picture provides a satisfactory description for the ground states of closed-shell nuclei. Practically, the self-consistent RHF equation can be derived from a variation of the energy functional
\begin{equation}\label{HF_equation}
	T_{ii'} + \sum^{A}_{j=1}\bar{V}_{iji'j}
	= \varepsilon_{i}\delta_{ii'},
\end{equation}
where $\varepsilon_{i}$ represents the single-particle energy, which excludes the rest mass of nucleon.
The one-body kinetic energy $T_{ii'}$ and the antisymmetric two-body interaction matrix elements $\bar{V}_{iji'j'}$ can be obtained as follows,
\begin{subequations}\label{TV}
	\begin{align}
		T_{ii'}   &= \int d\bm{x}\bar{\psi}_{i}(\bm{x}) (-i\bm{\gamma}\cdot\bm{\nabla}+M) \psi_{i'}(\bm{x}),\\
		\bar{V}_{iji'j'} &= \int d\bm{x}d\bm{x'} \bar{\psi}_{i}(\bm{x})\bar{\psi}_{j}(\bm{x'}) \Gamma_{\phi}D_{\phi}    \psi_{i'}(\bm{x})\psi_{j'}(\bm{x'}) \nonumber\\
		& - \int d\bm{x}d\bm{x'} \bar{\psi}_{i}(\bm{x})\bar{\psi}_{j}(\bm{x'}) \Gamma_{\phi}D_{\phi}
		\psi_{j'}(\bm{x})\psi_{i'}(\bm{x'}).
	\end{align}
\end{subequations}
where $\Gamma_{\phi}$ and $D_{\phi}$ denote the interaction vertex and meson (photon) propagators, respectively \cite{Bouyssy1987PRC36, Long2010PRC81, Geng2022PRC105}.

To further take into account the collective correlations explicitly, large-scale configuration mixing is introduced in the shell model approach. Starting from a complete set of single-particle basis, such as the self-consistent HF basis defined by the solutions of Eq. (\ref{HF_equation}), the model space for the shell model, namely the $P$-space, can be constructed on top of a frozen core $|\text{core}\rangle$,
\begin{equation}\label{ModelSpace}
	P = \sum_{n}|\Psi_{n}\rangle\langle\Psi_{n}|,\quad |\Psi_{n}\rangle  = \prod^{N_{\text{val}}}_{i=1}     c^{\dagger}_{n_{i}}|\text{core}\rangle,
\end{equation}
where $N_{\text{val}}$ is the number of valence nucleons, and $|\text{core}\rangle$ corresponds to the wave function of a selected closed-shell core. It shall be stressed that the creation operation ($c^\dagger$) is restricted within the valence space, and the outside space above it is unoccupied. Correspondingly, one can define the operator $Q$ as the complement of $P$, namely $Q=1-P$. In general, the nuclear system in a full Hilbert space can be described with the ``bare'' Hamiltonian $H = H_{0} + V$, where $H_{0}$ is the unperturbed Hamiltonian, and $V$ contains the residual interactions. However, the effective Hamiltonian $H^{\text{eff}}$ has to be considered when addressing the eigenvalue problem in a truncated model space,
\begin{equation}\label{SMEQ}
	PH^{\text{eff}}P|\Psi\rangle
	= EP|\Psi\rangle.
\end{equation}

There are several widely used methods to derive the effective interaction, such as the Krenciglowa-Kuo (KK) \cite{Krenciglowa1974NPA235} and the Lee-Suzuki (LS) methods \cite{Suzuki1980PTP64}, which are actually the implementations of the folded diagram approach. It is worth mentioning that, in these methods, the model space is assumed to be degenerate, which is a relatively strong limitation, especially for nuclei far from the $\beta$-stability line. Recently, extended versions of the Krenciglowa-Kuo (EKK) and Lee-Suzuki (ELS) methods have been proposed \cite{Takayanagi2011NPA852, Tsunoda2014PRC89}, which can be applied to the non-degenerate case. In the EKK scheme, the effective Hamiltonian can be obtained by the following iterations,
\begin{align}\label{Heff}
	H^{\text{eff}}_{n} = H^{\text{BH}}(E_{0})
	+ \sum^{\infty}_{k=1}
	\frac{1}{k!}\frac{d^{k}\widehat{Q}(E_{0})}{dE^{k}_{0}}
	\left\{H^{\text{eff}}_{n-1}-E_{0}\right\}^{k},
\end{align}
where the index $n$ denotes the step of the iterations, and the Bloch-Horowitz Hamiltonian satisfies $H^{\text{BH}}(E) = PH_{0}P + \widehat{Q}(E)$. The so-called $\widehat{Q}$-box is defined as,
\begin{align}
	\widehat Q(E) = & PVP + PV\frac{Q}{E-QHQ}VP.
\end{align}
Notably, Eq. (\ref{Heff}) can be interpreted as a Taylor series expansion of $H^{\text{eff}}$ around the starting energy $E_{0}$. Therefore, although the $\widehat{Q}$-box and the folded terms, specifically the sum on the right side of Eq. (\ref{Heff}), depend on the $E_{0}$, the total effective Hamiltonian $H^{\text{eff}}$ is independent of the starting energy.

For a single-shell valence space, where the unperturbed Hamiltonian can be considered degenerate, the effective interaction $V^{\text{eff}}$ in the KK scheme can be obtained from Eq. (\ref{Heff}) by assuming $E_{0}P=PH_{0}P$,
\begin{align}\label{Veff}
	V^{\text{eff}}_{n} = \widehat{Q}(E_{0})
	+ \sum^{\infty}_{k=1}
	\frac{1}{k!}\frac{d^{k}\widehat{Q}(E_{0})}{dE^{k}_{0}}
	\left\{V^{\text{eff}}_{n-1}\right\}^{k}.
\end{align}

In practical calculations, one cannot evaluate the $\widehat{Q}$-box exactly. However, it can be further expressed as an expansion around $H_{0}$:
\begin{align}
	\widehat{Q}(E)
	&= PVP + PV\frac{Q}{E-H_{0}}VP
	\nonumber\\
	&+ PV\frac{Q}{E-H_{0}}V\frac{Q}{E-H_{0}}VP + \cdots.
\end{align}
For better understanding, Fig. {\ref{Fig:Qbox}} shows the valence-linked $\widehat Q$-box diagrams, which contain the first-order $\widehat{Q}$-box in $V$, and the second-order core-polarization diagrams induced by one and two particle-hole ($p$-$h$) excitations, namely the $V^{\text{3p1h}}$ and $V^{\text{4p2h}}$ corrections. At an early stage, the $G$-matrix augmented with the second-order core-polarization corrections served as a satisfactory effective interaction for shell model calculations \cite{Kuo1966NP85, Kuo1968NPA114}. After the initial success, Barrett and Kirson found that large third-order terms tend to cancel the second-order contribution, indicating a lack of order-by-order convergence in this perturbation expansion \cite{Barrett1970NPA148}. However, when the summation of the folded diagrams extends to arbitrary orders as illustrated in Eq. (\ref{Veff}), the differences between the effective interactions containing the $\widehat{Q}$-box correction up to second order and those up to third order are small \cite{Hjorth1995PR261}.

\begin{center}
	\centering
	\includegraphics[width=1.0\linewidth]{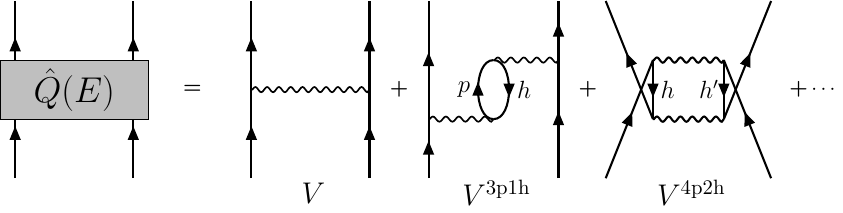}
	\figcaption{ Valence-linked $\widehat{Q}$-box diagrams contributing to the effective two-body Hamiltonian.}
	\label{Fig:Qbox}
\end{center}

In this paper, we consider the $\widehat{Q}$-box corrections up to the second order, in which the contributions of ladder diagrams are not included to avoid double-counting \cite{Vautherin1972PRC5, Shen2019PPNP109}. The folded terms with $k\ge10$ are terminated since the derivatives divided by the factorials are sufficiently small. Moreover, the core and single-particle energies are calculated consistently and all the corrections to the monopole part are neglected to keep the description of ground states of closed-shell nuclei almost unchanged. It is noteworthy that the rearrangement terms, induced by the density dependencies of the meson-nucleon coupling strengths, should be taken into account in deriving the effective Hamiltonian. The detailed derivations are presented in Appendix \ref{APPENDIX}.

It is worth emphasizing that all the CI-RHF calculations are carried out in a self-consistent manner. For a typical calculation, we outline the successive steps as follows. Firstly, starting from an empirical but general Lagrangian (\ref{Lagrangian}), spherical relativistic Hartree-Fock (RHF) calculations can be initially performed for specific nucleus. Then, aiming at interested observables, one can construct a suitable model space (\ref{ModelSpace}) by referring to the single-particle structure. Based on the RHF single-particle basis and the given Lagrangian, the core and single-particle energies are calculated consistently, then the effective Hamiltonian (\ref{Heff}) can be derived using the folded-diagram approach. Finally, by diagonalizing the effective Hamiltonian in the model space, namely solving Eq. (\ref{SMEQ}), one can obtain the properties of both the ground state and the low-lying states of the nucleus.

\section{RESULTS}\label{RESULTS}
As the first applications of the CI-RHF model, we focus on the ground state and low-lying excitation properties of light and medium-mass nuclei. In the subsequent calculations, we utilize different Lagrangians, specifically PKA1 \cite{Long2007PRC76} and PKO2 \cite{Long2008EPL82}. The PKO2 contains the degrees of freedom associated with the $\sigma$-scalar ($\sigma$-S), $\omega$-vector ($\omega$-V), $\rho$-vector ($\rho$-V) and photon $A$-vector ($A$-V) couplings, while PKA1 further considers $\pi$-pseudo-vector ($\pi$-PV) and $\rho$-tensor ($\rho$-T) couplings. The single-particle basis is provided by the spherical RHF calculations. Furthermore, the valence space for even-even nuclei $^{18}$O and $^{18\sim 28}$Ne is restricted to the $sd$ shell, composed of the $1d_{5/2}$, $2s_{1/2}$ and $1d_{3/2}$ orbits. In the current work, the excitations of the Dirac sea are negligible due to the large energy denominator. Additionally, the BIGSTICK code is used for the diagonalization of shell model Hamiltonians \cite{Johnson2018arxiv}.

\subsection{Convergence of intermediate-state excitations}
In general, one needs about 10 oscillator quanta excitations to obtain fully converged effective TBMEs \cite{Shurpin1977NPA293,Sommermann1981PRC23, Engeland2014NPA928}. However, in this work, the effective TBMEs are calculated with the self-consistent Hartree-Fock single-particle basis. Before realistic applications, it is necessary to examine the convergence of intermediate-state excitations, namely, to determine the energy truncation of particle states ($p$).

Using the Lagrangian PKA1 and considering the simple nucleus $^{18}$O as an example, we calculate the ``bare'' TBMEs, namely the first-order $\widehat{Q}$-box diagram, and the effective TBMEs with different energy truncations. As shown in Fig. {\ref{Fig:Veff_conv}}, the TBMEs are indicated by the red dots. The vertical coordinate corresponds to the results obtained with energy truncation of $\varepsilon \leqslant 80$ MeV, while the horizontal coordinates represent the ``bare'' TBMEs and the effective TBMEs with energy truncations of $\varepsilon \leqslant 20$, $40$, and $60$ MeV, for Figs. \ref{Fig:Veff_conv}(a-d), respectively. It is evident that the corrections induced by the core-polarization effects are significant, and the vast majority of these effects are captured when $\varepsilon \leqslant 20$ MeV, as illustrated in Figs.~\ref{Fig:Veff_conv}(a) and \ref{Fig:Veff_conv}(b). Furthermore, it can be observed that the calculated effective TBMEs tend to converge when $\varepsilon \leqslant 40$ MeV, and they remain almost unchanged from $\varepsilon \leqslant 60$ MeV to $\varepsilon \leqslant 80$ MeV. Therefore, considering the practical computational limitations, we consider all possible particle-hole excitations with particle states energy truncation of $\varepsilon \leqslant 80$ MeV for light and medium-mass nuclei. Additionally, it should be noted that the effective TBMEs given by the Lagrangian PKO2 are also converged under such truncation.

\begin{center}
	\includegraphics[width=1.0\linewidth]{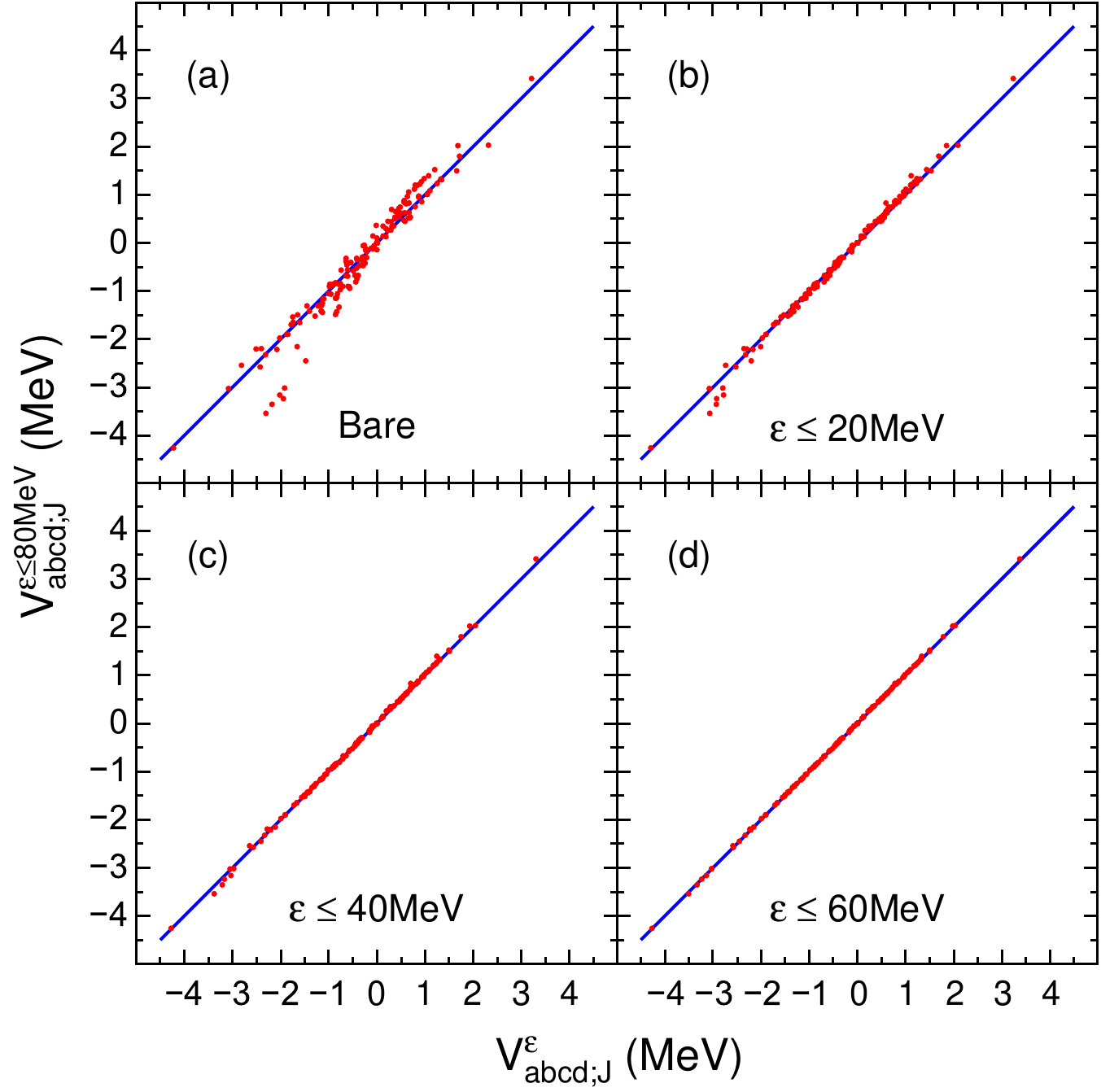}
	\figcaption{Comparison of two-body interaction matrix elements $V_{abcd;J}$ (MeV) of $^{18}$O obtained with different particle state energy truncations using the Lagrangian PKA1. The diagonal line indicates the equivalence of TBMEs within different energy truncations. See text for more explanations.}
	\label{Fig:Veff_conv}
\end{center}

\subsection{Core-polarization effects}
The core-polarization corrections were initially calculated by Bertsch for $^{18}$O and $^{42}$Sc \cite{Bertsch1965NP74}, and subsequently their effects were quantitatively evaluated in shell model calculations by Kuo and Brown \cite{Kuo1966NP85}. It has been shown that the dominant contributions of the core-polarization diagrams originate from the quadruple component, providing a microscopic interpretation for the empirical long-range quadruple force \cite{Brown1967NPA92}. Additionally, Shurpin and the collaborators found that the core-polarization corrections are significantly influenced by the single-particle basis \cite{Shurpin1977NPA293}. As the first attempt, starting from the density-dependent Lagrangians PKA1 and PKO2, Fig. {\ref{Fig:Core_polar}} illustrates the yrast band structure of $^{18}$O calculated with the self-consistent Hartree-Fock basis. In contrast to the calculations using ``bare'' PKA1 and PKO2 interactions, which present rather compressed spectra, fairly good agreement with the experimental data is achieved after considering the core-polarization corrections.

\begin{center}
	\includegraphics[width=1.0\linewidth]{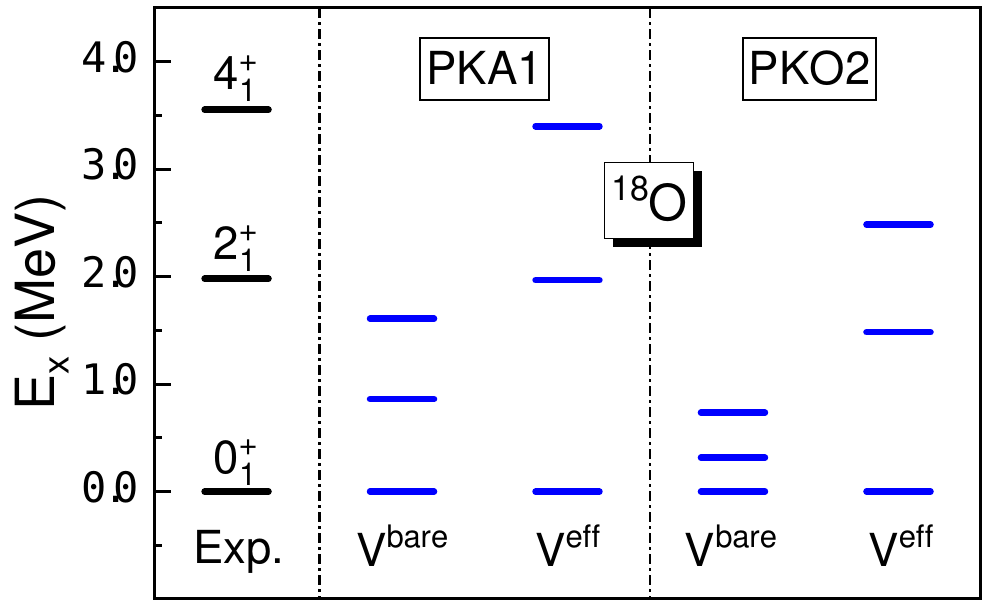}
	\figcaption{The calculated yrast bands of $^{18}$O based on Lagrangians PKA1 and PKO2, compared with experimental results from \cite{NNDC}. Core-polarization corrections are not included in the case $V^{\text{bare}}$. Note that the energy of the ground state ($0^{+}_{1}$) is set to zero.}
	\label{Fig:Core_polar}
\end{center}

In fact, our results are similar to those obtained using microscopic $G$-matrix with a harmonic-oscillator basis \cite{Kuo1966NP85, Hjorth1995PR261}, but not with the Brueckner Hartree-Fock basis \cite{Shurpin1977NPA293}. The latter typically yields reduced core-polarization corrections, due to the small radii for hole states and large ones for particle states, resulting in compressed spectra. Recently, a self-consistent Dirac Brueckner-Hartree-Fock approach has been implemented \cite{Shen2019PPNP109}, providing a better description of the binding energy and charge radii of $^{16}$O, which could improve this issue.

As pointed out by Brown and Kuo \cite{Brown1967NPA92}, the core-polarization corrections primarily arise from the long-range quadrupole component of neutron-proton isoscalar interactions. Consequently, for the stable nucleus $^{18}$O, it can be observed that the core-polarization corrections, which are associated with the quadrupole properties of nuclei, are nearly identical for different Lagrangians PKA1 and PKO2. 

However, the spectra obtained from the ``bare'' PKA1 and PKO2 exhibit notable differences. Specifically, these differences are determined by the isovector two-body matrix elements $V^{\text{bare}}_{aaaa;J}$, where index $a$ denotes the neutron orbit $\nu 1d_{5/2}$. As an illustration, we extracted the contributions from various meson couplings to these TBMEs, as shown in Table \ref{Tab:MS_TBME}. It is observed that the $\sigma$-scalar and $\omega$-vector couplings provide crucial contributions to these TBMEs as described by PKO2, but less for PKA1. This is due the fact that the $\rho$-tensor coupling in PKA1 presents notable contributions, which results in distinct differences in the spectra obtained from ``bare'' PKA1 and PKO2.

\begin{center}
    \tabcaption{The calculated ``bare'' two-body interaction matrix elements $V^{\text{bare}}_{aaaa;J}$ (MeV) of $^{18}$O obtained using the Lagrangians PKA1 and PKO2. Here the index $a$ represents the neutron orbit $\nu1d_{5/2}$. The contributions from different meson couplings are also presented. See the text for more explanations.}  \label{Tab:MS_TBME}
    \begin{adjustbox}{max width=\columnwidth}
    \begin{tabular}[width=1.0\linewidth]{ccccccc}
        \hline\hline
        & J & V$^{\text{bare}}_{aaaa;J}$ & $\sigma$-S+$\omega$-V & $\rho$-V & $\rho$-T & $\pi$-PV \\
        \hline \multirow{3}{*}{PKA1}
        & 0 & -1.92 & -0.68 & 0.92 & -2.13 & -0.03 \\
        & 2 & -1.13 & -0.82 & 0.24 & -0.48 & -0.07 \\
        & 4 & -0.63 & -0.56 & 0.11 & -0.21 & 0.03  \\
        \hline \multirow{3}{*}{PKO2}
        & 0 & -1.35 & -2.48 & 1.13 & 0.00  & 0.00  \\
        & 2 & -1.02 & -1.32 & 0.29 & 0.00  & 0.00  \\
        & 4 & -0.66 & -0.79 & 0.13 & 0.00  & 0.00  \\
        \hline\hline
    \end{tabular}
    \end{adjustbox}
\end{center}

\subsection{Description of doubly magic nuclei}
Due to the presence of the significant shell gaps, the ground state properties of doubly magic nuclei are frequently employed to constrain phenomenological Lagrangians within the mean field framework, such as PKA1 and PKO2. After considering the core polarization corrections and configuration mixing, the beyond-mean-field effects can be contained explicitly in the CI-RHF calculations. To elucidate the distinctions between the RHF and CI-RHF calculations, we present the calculated ground state binding energies and charge radii of doubly magic nuclei $^{16}$O, $^{40}$Ca and $^{48}$Ca in Table \ref{Tab:BE_RC}.
It should be noted that in the CI-RHF calculations, the valence space for $^{16}$O consists of the $p$ shell and $sd$ shell, while for $^{40}$Ca, it includes the $2s_{1/2}$, $1d_{3/2}$, $1f_{7/2}$, and $2p_{3/2}$ orbitals. Additionally, we adopt the $pf$ shell as valence space for $^{48}$Ca.

It is observed that both the binding energies and charge radii calculated using the CI-RHF model (denoted as CI in Table \ref{Tab:BE_RC}) are larger than those given by the RHF calculations (denoted as HF in Table \ref{Tab:BE_RC}). For PKO2, the RHF results underestimate both the binding energies and the charge radii, as compared to the experimental values. However, after explicitly incorporating beyond-mean-field effects, the CI-RHF calculations improve the descriptions of $^{16}$O, $^{40}$Ca, and $^{48}$Ca. Conversely, for the Lagrangian PKA1, the CI-RHF results for these nuclei are slightly overestimated, whereas the RHF results align more closely with experimental data. Even though, the beyond-mean-field corrections to the ground state bulk properties of doubly magic nuclei are relatively small, which indicates that the phenomenological Lagrangians PKA1 and PKO2 can serve as reasonable inputs for the CI-RHF calculations.

\begin{center}
    \centering
    \tabcaption{The calculated binding energies and charge radii of $^{16}$O, $^{40}$Ca and $^{48}$Ca using the RHF and CI-RHF model with Lagrangians PKA1 and PKO2, compared to the experimental values (taken from \cite{Angeli2013ADNDT, Wang2017CPC41}). }
    \label{Tab:BE_RC}
    \setlength{\tabcolsep}{4pt}\renewcommand{\arraystretch}{1.25}
    \begin{adjustbox}{max width=\columnwidth}
    \begin{tabular}[width=1.0\linewidth]{lcccccc}
        \hline\hline
        \multicolumn{1}{c}{}&\multicolumn{3}{c}{Binding Energy (MeV)}&\multicolumn{3}{c}{Charge radii (fm)} \\
                    & $^{16}$O  & $^{40}$Ca & $^{48}$Ca & $^{16}$O  & $^{40}$Ca & $^{48}$Ca \\\hline
        Exp.        & 127.62    & 342.05    & 416.00    & 2.699     & 3.478     & 3.477     \\\hline
        PKA1(HF)    & 127.16    & 341.48    & 416.04    & 2.802     & 3.530     & 3.495     \\
        PKA1(CI)    & 128.41    & 343.08    & 416.95    & 2.860     & 3.535     & 3.495     \\ \hline
        PKO2(HF)    & 127.09    & 340.70    & 415.25    & 2.673     & 3.434     & 3.457     \\
        PKO2(CI)    & 127.60    & 341.83    & 415.85    & 2.715     & 3.439     & 3.457     \\
        \hline\hline
    \end{tabular}
    \end{adjustbox}
\end{center}

\subsection{Description of Neon isotopes}
In general, with the addition of more valence particles, it is necessary to consider the mass dependence of the harmonic oscillator basis \cite{Wildenthal1984PPNP11} and the $G$-matrix \cite{Hosaka1985NPA444}, as well as the effects of the three-body force \cite{Stroberg2017PRL118}. For instance, an empirical mass dependence $(18/A)^{0.3}$ is assumed for the USD Hamiltonians \cite{Brown2006PRC74}. Within the CI-RHF framework, it is expected that the initial Hartree-Fock calculations can capture the effect of changing mean field, while the complicated many-body effects can be evaluated through the density-dependence of the meson-nucleon coupling strength \cite{Long2006PLB640, Long2008EPL82, Long2007PRC76}. Consequently, for the nuclei with more than two valence nucleons, no additional parameters are introduced in the CI-RHF calculations.

\begin{center}
    \centering
    \tabcaption{The calculated binding energies (in MeV) for the even-even $^{18\sim28}$Ne isotopes using the CI-RHF model with Lagrangians PKA1 and PKO2, compared with the experimental values (taken from \cite{Wang2017CPC41}). }
    \label{Tab:Ne_mass}
    \begin{adjustbox}{max width=\columnwidth}
    \begin{tabular}[width=1.0\linewidth]{lcccccc}\hline\hline
                & $^{18}$Ne & $^{20}$Ne & $^{22}$Ne & $^{24}$Ne & $^{26}$Ne & $^{28}$Ne \\\hline
        Exp.    & 132.14    & 160.65    & 177.77    & 191.84    & 201.55    & 206.86    \\
        PKA1    & 133.07    & 163.45    & 181.04    & 194.25    & 204.80    & 209.83    \\
        PKO2    & 131.87    & 160.67    & 179.99    & 193.99    & 201.80    & 206.56    \\
        \hline\hline
    \end{tabular}
    \end{adjustbox}
\end{center}

The neon isotopes span the entire neutron $sd$ shell, provide ideal examples for studying low-lying spectra and their isotopic evolution. In the following, we first restrict the valence space to the $sd$ shell and present the results for the even-even isotopes $^{18\sim 28}$Ne.

As shown in Table \ref{Tab:Ne_mass}, the binding energies of neon isotopes are described reasonably by the CI-RHF calculations, despite a slight overestimation by the Lagrangian PKA1. These results are consistent with the behavior observed for doubly magic nuclei, where the Lagrangian PKA1 similarly overestimates the binding energies, while the Lagrangian PKO2 exhibits better agreement with experimental data. On the other hand, the CI-RHF calculations provide a reasonable description of the isotopic shift of charge radii for neon isotopes, despite the values being systematically underestimated by PKO2, as illustrated in Fig. \ref{Fig:Ne_Rc}. Considering that the $\pi$-pseudo-vector and $\rho$-tensor couplings can enhance the quadruple correlations and give rise to more notable deformation, as indicated by previous studies \cite{Geng2020PRC101, Geng2022PRC105}, it is not difficult to understand that PKA1 yields larger charge radii than PKO2 for the ground states of neon isotopes. Moreover, it can be observed that $^{20}$Ne presents the largest charge radius. This is understandable because $^{20}$Ne is strongly deformed, leading to a significant occupation of protons (occupancy number 0.62 and 0.42 for PKA1 and PKO2, respectively) in the more extended orbits $\pi2s_{1/2}$ and $\pi1d_{3/2}$. Notably, the charge radius of $^{24}$Ne is significantly smaller compared to that of neighboring isotopes, which may serve as an indicator of the neutron subshell $N = 14$.

\begin{center}
	\includegraphics[width=1.0\linewidth]{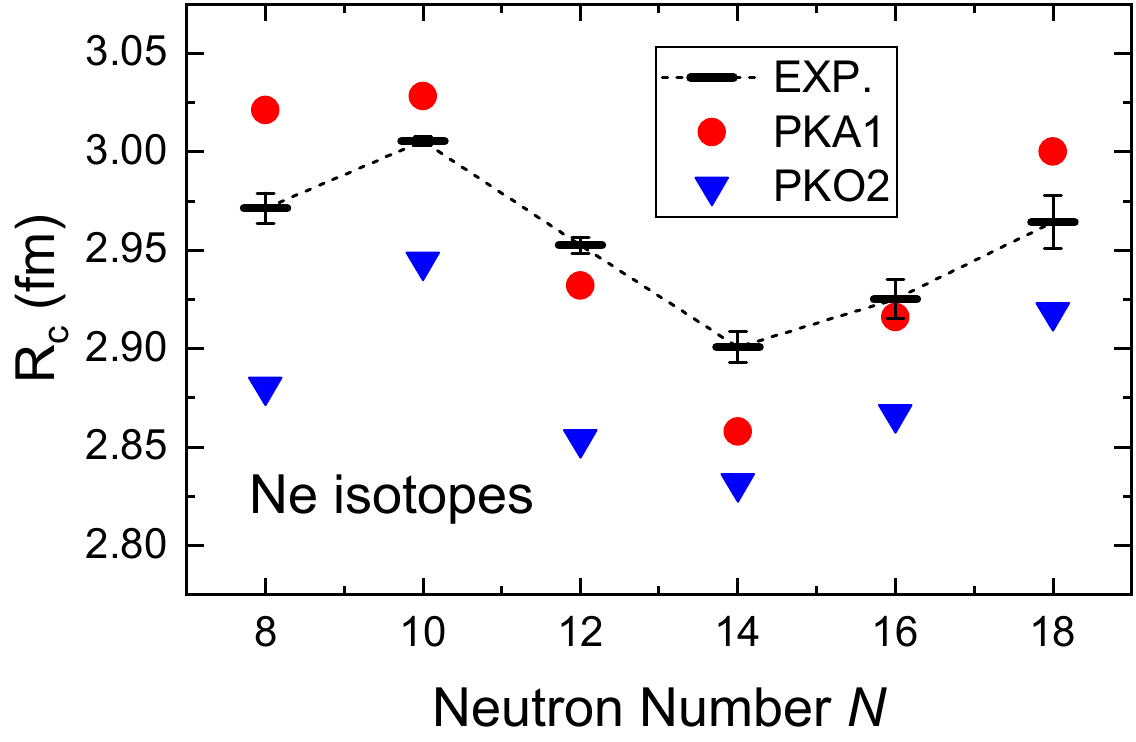}
	\figcaption{The calculated charge radii $R_{\text{c}}$ of neon isotopes using the CI-RHF model with Lagrangians PKA1 and PKO2, compared with experimental data. The dashed line represents the trend of experimental values (taken from \cite{Angeli2013ADNDT}).}
	\label{Fig:Ne_Rc}
\end{center}


\begin{figure*}[htbp]
	\centering
	\includegraphics[width=0.85\linewidth]{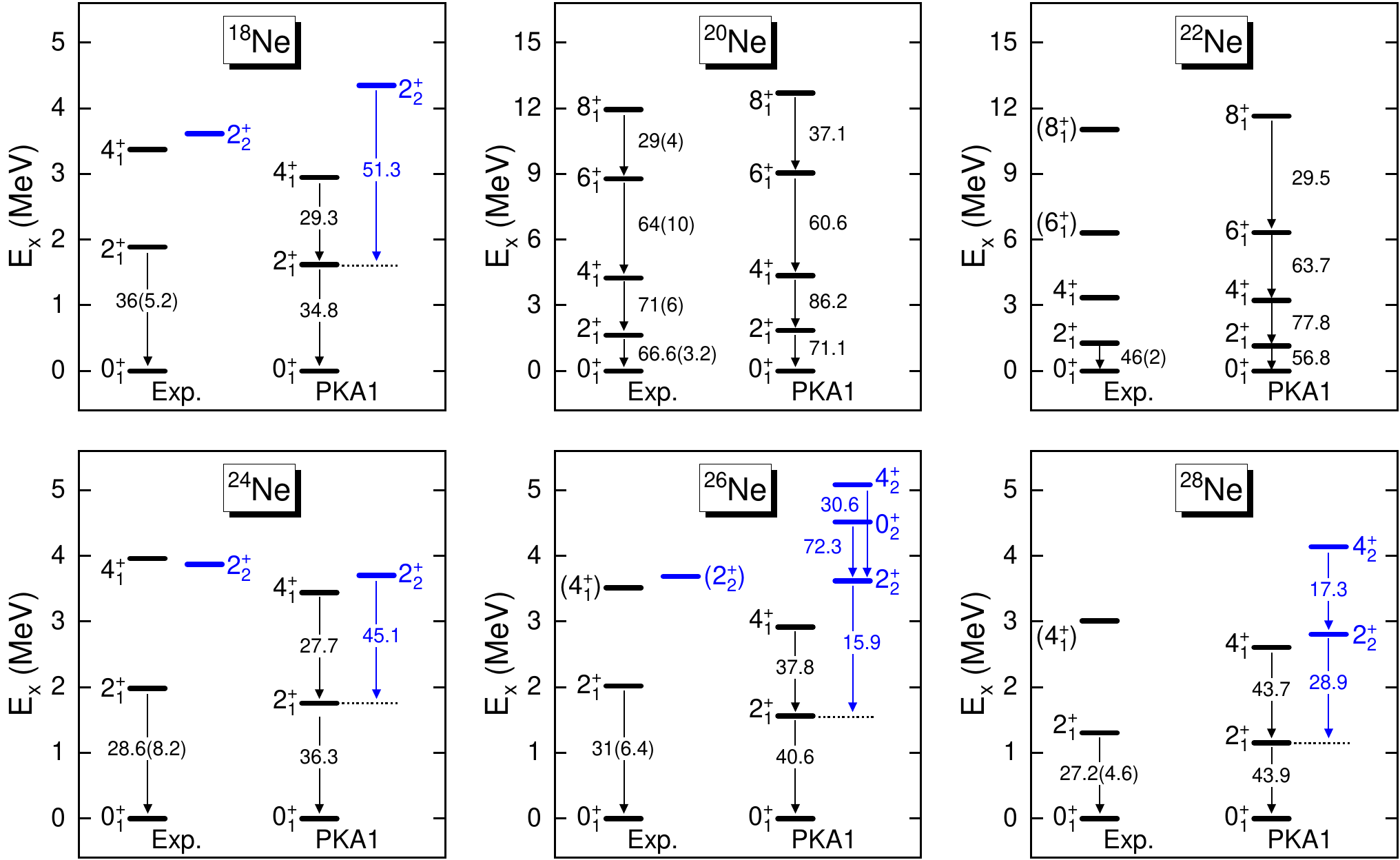}
	\figcaption{Comparison of low-lying spectra of neon isotopes between experimental data \cite{NNDC,Pritychenko2016ADNDT107} and theoretical results obtained using the CI-RHF model with Lagrangian PKA1. The available reduced transition probabilities $B(E2)$ (in e$^{2}$fm$^{4}$) within and between bands are also shown. Note that the energies of the ground states ($0^{+}_{1}$) are renormalized to zero.}
	\label{Fig:Ne_spec}
\end{figure*}

Besides the ground state properties, the CI-RHF model also aims at the low-lying excitations using the same Lagrangian. Taking PKA1 as the candidate, Fig. {\ref{Fig:Ne_spec}} further shows the low-lying spectra of even-even isotopes $^{18\sim 28}$Ne. The reduced electric quadruple transition probabilities from the initial state $\Psi_{i}$ to the final state $\Psi_{f}$ are calculated as follows,
\begin{align}
	B(E2; \Psi_{i} \rightarrow \Psi_{f})
	= \frac{1}{2J_{i} + 1}
	\left| \langle \Psi_{f} || e_{\tau_z} r^2 Y_2 || \Psi_{i} \rangle \right|^2
\end{align}
It should be noted that $e_{\tau_z}$ denotes the bare charge for protons or neutrons in the full space, while effective charges must be introduced in a truncated shell model space to account for the 2$\hbar\omega$ core excitations. In fact, a reasonable quantitative description of $B(E2)$ would require the inclusion of higher-order corrections to effective charge \cite{Siegel1970NPA145,Stroberg2019ARNS69}. Considering the limitation of practical calculations, in the current work, we use empirical effective charges of 1.29 and 0.49 for protons and neutrons, respectively, which are averages over orbits and masses in the $sd$ shell obtained through a least-squares fit \cite{Brown1988ARNPS38}.

As illustrated in Fig. {\ref{Fig:Ne_spec}}, it is clear that the calculated low-lying spectra are in good agreement with the experimental data for both the yrast and non-yrast bands. Specifically, the excited energies for $^{20,22}$Ne are slightly overestimated by the present calculation, while it yields a relatively compressed spectrum for $^{18}$Ne and even-even isotopes $^{24\sim 28}$Ne. The trend of the reduced transition probabilities $B(E2; 2^{+}_{1} \rightarrow 0^{+}_{1})$ is consistent with the experimental results, despite the value for $^{28}$Ne being overestimated. In fact, earlier experimental measurement yielded a $B(E2)$ with larger central value (53.8$\pm$27.2 e$^{2}$fm$^{4}$) for $^{28}$Ne \cite{Pritychenko1999PLB461}, as given by Monte Carlo shell model calculation \cite{Utsuno1999PRC60}. Such inconsistency remains unexplained. On the other hand, it should be noted that the effective charges are orbit-dependent and can be influenced by the single-particle wave functions of the specific nuclide. As one moves away from stable nuclei toward neutron-rich side, the effective charges typically decrease \cite{Sagawa2001PRC63}. Consequently, aiming at reliable analysis and prediction for transition properties, it is necessary to calculate the effective charges with the self-consistent Hartree-Fock single-particle wave functions in the future.

Furthermore, it can be observed that both $^{24}$Ne and $^{26}$Ne exhibit higher excited state energies $E(2^{+}_{1})$ compared to their neighboring isotopes. In general, large shell gaps result in relatively high excited state energies $E(2^{+}_{1})$ and low quadrupole collectivity. Along the isotopic chain, the $B(E2; 2^{+}_{1} \rightarrow 0^{+}_{1})$ values increase from $^{18}$Ne to $^{20}$Ne and $^{22}$Ne but are significantly reduced at $N=14$ and $N=16$. $^{20}$Ne and $^{22}$Ne are strongly deformed \cite{NNDC,Ebran2018PRC97}, with a similar energy ratio $E(4^{+}_{1})/E(2^{+}_{1})$ of approximately 2.7, both experimentally and theoretically. However, the ratios $E(4^{+}_{1})/E(2^{+}_{1})$ for $^{24}$Ne and $^{26}$Ne are close to 2.0, indicating near pure collectivity of vibration. Therefore, combined with the isotopic evolution of charge radii shown in the lower panel of Fig. {\ref{Fig:Ne_Rc}}, a notable subshell closure at $N=14$ is indicated in neon isotopes, as well as a less pronounced one at $N=16$.

As discussed above, the CI-RHF calculations provide a reasonable description of even-even $^{18\sim28}$Ne nuclei when the valence space is restricted to the $sd$ shell. However, extensive studies on atomic masses \cite{Thibault1975PRC12, Chaudhuri2013PRC88}, charge radii \cite{Yordanov2012PRL108}, and excitation spectra \cite{Motobayashi1995PLB346, Yanagisawa2003PLB566, LiuHN2017PLB767} have revealed that the neutron magicity $N=20$ disappears for nuclei in the ``island of inversion'', indicating the intrusion of the $pf$ shell. Consequently, it becomes necessary to obtain a multi-shell Hamiltonian to account for the cross-shell excitations. For instance, the SDPF-M effective Hamiltonian was constructed to explain the anomalous properties of neutron-rich nuclei around $N=20$ \cite{Utsuno1999PRC60}, with the model space comprising the $sd$ shell and additional $1f_{7/2}$ and $2p_{3/2}$ orbits. More recently, taking the full $sd$ shell and neutron $pf$ shell as valence space, the effective Hamiltonian SDPF-U-MIX has been proposed \cite{Caurier2014PRC90}, to study a broad range of nuclear properties. Notably, experimental guidance on single-particle energies is typically lacking for nuclei with extreme $N/Z$ ratio, and it is challenging to determine the effective TBMEs in such a large model space by fitting experimental data.

In the CI-RHF framework, starting from the empirical but general Lagrangian, it is convenient to derive the multi-shell Hamiltonian by incorporating the EKK method \cite{Takayanagi2011NPA852, Tsunoda2014PRC89}, since no additional parameters are needed to be readjusted. Taking the neutron-rich nuclei $^{30}$Ne and $^{32}$Ne as examples, we performed the $sd$-$pf$ cross-shell calculations with the effective Hamiltonian derived from the Lagrangian PKA1, in comparison to the results given by the effective Hamiltonian SDPF-U-MIX. In order to reduce the dimension of configurations, the CI-RHF calculations with PKA1 include only the neutron $1f_{7/2}$ and $2p_{3/2}$ orbits in $pf$ shell as active orbits, in addition to the $sd$ shell for proton and neutron. For comparison, the left panel of Fig. {\ref{Fig:Ne_cross}} also show the theoretical results of $^{30}$Ne calculated with PKA1 and USDB Hamiltonians \cite{Brown2006PRC74}, for which the model space is restricted to the $sd$ shell.

\begin{center}
	\centering
	\includegraphics[width=1.0\linewidth]{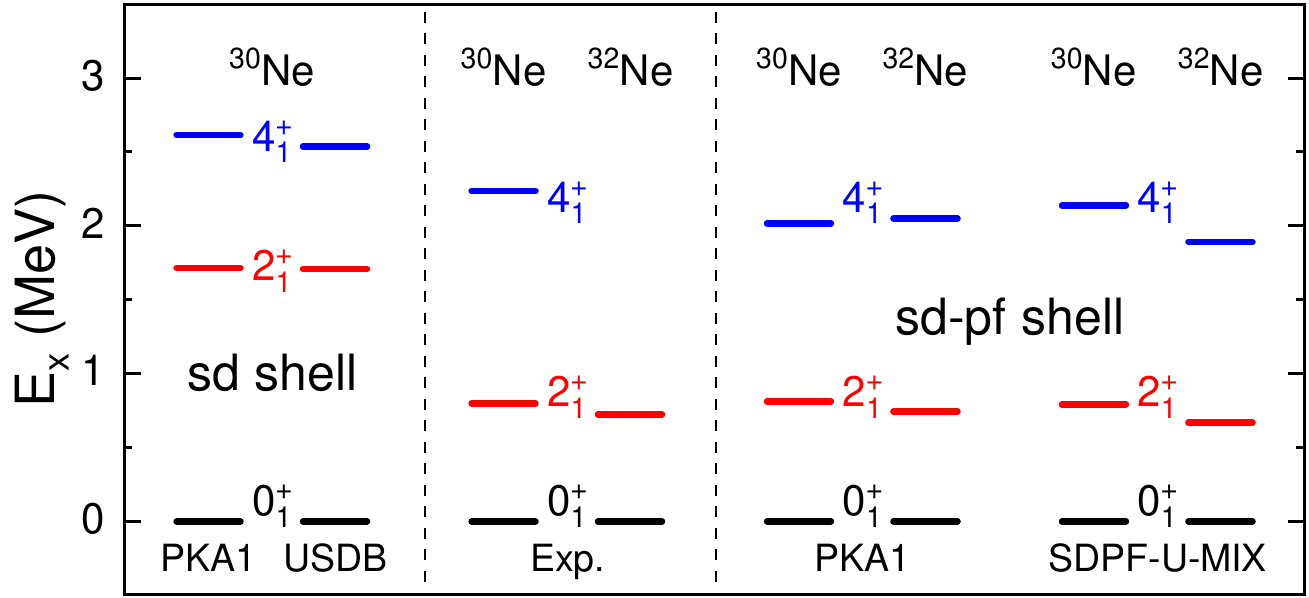}
	\figcaption{Cross-shell calculations of the excitation spectra for $^{30}$Ne and $^{32}$Ne, compared with experimental data \cite{NNDC, Yanagisawa2003PLB566, Doornenbal2009PRL103}. The left panel shows theoretical results with the valence space restricted to the $sd$ shell. Note that the energies of the ground states ($0^{+}_{1}$) have been renormalized to zero. See the text for more explanations.}
	\label{Fig:Ne_cross}
\end{center}

It is observed that the excited energies $E(2^{+}_{1})$ of $^{30}$Ne are significantly overestimated with the neutron closed-shell configuration for both the PKA1 and USDB Hamiltonians. After considering the neutron cross-shell excitations, the CI-RHF calculations using the same Lagrangian PKA1 successfully reproduced the low-lying spectra, yielding results comparable to those obtained with the SDPF-U-MIX Hamiltonian \cite{Caurier2014PRC90}. It is noteworthy that the effective single-particle energy of the neutron $2p_{3/2}$ orbit, obtained with Lagrangian PKA1, is approximately 1 MeV lower than that of the $1f_{7/2}$ orbit, in contrast to the conventional spherical shell ordering. In fact, recent cross sections measurements for Ne isotopes also suggest that the valence neutrons in $^{31}\text{Ne}$ occupy low-$l$ orbits \cite{Nakamura2009PRL103, Takechi2012PLB707}, leading to a halo structure in this nucleus \cite{Nakamura2014PRL112}. It would be intriguing to investigate the mechanism driving shell evolution and halo formation around the $N=20$ ``island of inversion'' in future studies.

\section{CONCLUSION} \label{CONCLUSION}
In this paper, we have proposed a method to consistently construct the configuration interactions for shell model calculations based on relativistic Hartree-Fock (RHF) theory, which is referred as the CI-RHF model. In the CI-RHF framework, initial RHF calculations capture the effect of changing mean field, and the density-dependent meson-nucleon coupling allows for a consistent treatment of complex many-body effects. Consequently, the core and single-particle energies, as well as the effective two-body matrix elements (TBMEs), can be derived from the same Lagrangian for each nuclide, which provides a robust solution for determining the effective Hamiltonian for any given model space. Based on that, the CI-RHF model is applicable for studying the properties of a wide range of nuclei with a few well-defined parameters in the phenomenological Lagrangian.

Taking the simple nucleus $^{18}$O as an example, we have checked the convergence of intermediate-state excitations in evaluating the effective interactions and studied the core-polarization effects, which are crucial for reproducing the structure of the yrast band. Furthermore, we have shown that the CI-RHF model provide a satisfactory description of both the bulk properties and low-lying excited spectra for doubly magic nuclei and even-even nuclei $^{18\sim 28}$Ne. The isotopic evolution characteristics of charge radii and low-lying spectra indicate a notable subshell at $N=14$ in the middle of the $sd$ shell for neon isotopes. Finally, by taking into account neutron cross-shell excitations from $sd$ shell orbits to the $1f_{7/2}$ and $2p_{3/2}$ orbits, the CI-RHF calculations successfully reproduced the low-lying spectra for more neutron-rich nuclei $^{30}$Ne and $^{32}$Ne without readjusting any parameters. These promising results demonstrate that the CI-RHF model can provide equally reasonable descriptions of both stable and unstable nuclei using the same Lagrangian, which is crucial for making reliable predictions for nuclei near the proton or neutron dripline.

It should be noted that the CI-RHF calculations remain limited due to computational restrictions on the diagonalization of the shell model Hamiltonians, similar as met in the conventional CISM calculations. It would be interesting to diagonalize the shell model Hamiltonians in the nonorthogonal deformed basis by incorporating the Generator Coordinate Method \cite{Dao2022PRC105,Griffin1957PR108,Peierls1957PPS70}, which could help us reach regions beyond the limit of the CISM and achieve a unified description for nuclei throughout the entire nuclide chart.

\end{multicols}

\appendix
\section{DETAILS OF EFFECTIVE HAMILTONIAN CALCULATIONS}\label{APPENDIX}
\subsection{Rearrangement effects for the density-dependent Hamiltonian}
To account for the complicated in-medium effects in nuclear many-body system, the density dependence is typically introduced into phenomenological Lagrangians or nuclear forces. For instance, the meson-nucleon coupling strengths depend on the one-body density in the density-dependent relativistic Hartree-Fock model \cite{Long2006PLB640}. In the following, we consider a general density-dependent Hamiltonian to describe the many-body system in a full Hilbert space:
\begin{align}\label{Ham_gen}
	H = \sum_{ii'}T_{ii'}c^{\dagger}_{i}c_{i'} 
	+ \frac{1}{4}\sum_{iji'j'}\bar{V}_{iji'j'}(\rho)c^{\dagger}_{i}c^{\dagger}_{j}c_{j'}c_{i'},
\end{align}
where the kinetic energy matrices $T_{ii'}$ and antisymmetric two-body interaction matrices $\bar{V}_{iji'j'}$ are expressed as Eq. (\ref{TV}) in the relativistic framework. Considering a general many-body wave function $|\Psi\rangle$, the one-body and two-body nucleon density matrices can be obtained as follows:
\begin{align}
	\rho_{ii'} = \langle\Psi|c^{\dagger}_{i}c_{i'}|\Psi\rangle,
	\qquad
	\rho_{iji'j'} = \langle \Psi|c^{\dagger}_{i}c^{\dagger}_{j}c_{j'}c_{i'}|\Psi\rangle.
\end{align}
Then one may express the energy functional $E$ with respect to the one-body and two-body density matrices as a Taylor series expansion around the starting one-body density $\rho^{0}\equiv\{\rho^{0}_{11},\rho^{0}_{12},\cdots\}$:
\begin{align}\label{EDFRHO}
	E 
	&= \sum_{ii'} T_{ii'}\rho_{ii'}
	+ \frac{1}{4}\sum_{iji'j'}
	\bar{V}_{iji'j'}(\rho^{0})\rho_{iji'j'}
	+ \frac{1}{4}\sum_{iji'j'}\sum_{kl}
	\bar{V}^{'(kl)}_{iji'j'}(\rho^{0})\rho_{iji'j'}
	(\rho_{kl}-\rho^{0}_{kl})
	+ \cdots,
\end{align}
where $\bar{V}^{'(kl)}$ represents the first-order derivative with respect to the one-body nucleon density matrix $\rho_{kl}$. It is clear that, besides the normal terms, namely the linear terms of the density matrices $\rho_{ii'}$ or $\rho_{iji'j'}$, there are higher-order terms induced by the rearrangement effects in the energy functional (\ref{EDFRHO}). It should be emphasized that, before the variation of energy functional, one cannot evaluate the rearrangement effects exactly, which include all the terms up to infinite order.

In practical calculations, to accelerate the convergence of the Taylor series expansion, the one-body density of the Hartree-Fock ground state is adopted as the starting point of the expansion, namely $\rho^{0} = \rho^{\text{HF}}$. On the other hand, for all nonlinear terms, one can approximate the one-body or two-body nucleon density matrices with $\rho^{0}_{ij}$ or $\rho^{0}_{iji'j'}$ to ensure the linear density-dependence of the energy functional. Consequently, the energy functional is approximated as follows:
\begin{align}\label{EDF}
	E\simeq E^{\text{re}}_{0}
	+ \sum_{ii'} T^{\text{re}}_{ii'}\rho_{ii'}
	+ \frac{1}{4}\sum_{iji'j'}
	\bar{V}_{iji'j'}(\rho^{\text{HF}})\rho_{iji'j'},
\end{align}
where the zero-order rearrangement term $E^{\text{re}}_{0}$ and kinetic energies containing the rearrangement term corrections $T^{\text{re}}_{ii'}$ are defined as:
\begin{align}\label{Rerrange}
	E^{\text{re}}_{0}
	=
	-\frac{1}{4}\sum_{iji'j'}\sum_{kl}
	\bar{V}^{'(kl)}_{iji'j'}(\rho^{\text{HF}})
	\rho^{\text{HF}}_{iji'j'}\rho^{\text{HF}}_{kl},
	\qquad
	T^{\text{re}}_{ii'}
	=
	T_{ii'}+\frac{1}{4}\sum_{kjlj'}
	\bar{V}^{'(ii')}_{kjlj'}(\rho^{\text{HF}})
	\rho^{\text{HF}}_{kjlj'}.
\end{align}
It is clear that, in this case, only the monopole part of the Hamiltonian is modified by the rearrangement terms. In the future, one may evaluate the rearrangement effects more precisely after the variation of the energy functional.

\subsection{The derivations of effective Hamiltonian for shell model}
The energy eigenvalue equation can be obtained from a variation of the energy functional (\ref{EDF}) with respect to the coefficients $C_{n}$, namely the probability amplitude of a specific basis $\Psi_{n}$. Considering that the shell model calculations are typically restricted to a truncated model space (\ref{ModelSpace}), the effective Hamiltonian must be introduced to account for the configurations beyond the model space,
\begin{align} \label{Hmn}
	\sum_{n}H^{\text{eff}}_{mn}C_{n} = EC_{m},
	\qquad
	H^{\text{eff}}_{mn}
	= E^{\text{eff}}_{\text{c}}\delta_{mn}
	+ \sum_{ii'}\varepsilon^{\text{eff}}_{ii'}\rho^{mn}_{ii'}
	+ \frac{1}{4}\sum_{iji'j'}\bar{V}^{\text{eff}}_{iji'j'}
	\rho^{mn}_{iji'j'},
\end{align}
where the one-body and two-body nucleon density matrices from the initial state $\Psi_{n}$ to final state $\Psi_{m}$ are defined as:
\begin{align}
	\rho^{mn}_{ii'}
	= \langle\Psi_{m}|c^{\dagger}_{i}c_{i'}|\Psi_{n}\rangle,
	\qquad
	\rho^{mn}_{iji'j'}
	= \langle \Psi_{m}|c^{\dagger}_{i}c^{\dagger}_{j}c_{j'}c_{i'}|\Psi_{n}\rangle.
\end{align}
It is clear that the input for shell model calculations is the effective Hamiltonian matrix elements $H^{\text{eff}}_{mn}$, which include the core and single-particle parts as well as the residual two-body interaction matrix elements. For conventional shell model, the core energy $E^{\text{eff}}_{\text{c}}$ is taken from experimental data, and the single-particle energies $\varepsilon^{\text{eff}}_{ii'}$ are typically determined by fitting or experimental data.

In this work, starting from a general Lagrangian (\ref{Lagrangian}), both the core and single-particle energies, along with the rearrangement terms, are calculated consistently based on the Hartree-Fock single-particle basis:
\begin{align}
	E^{\text{eff}}_{\text{c}}
	= \sum^{N_{\text{c}}}_{i=1}T^{\text{re}}_{ii} 
	+ \frac{1}{2}\sum^{N_{\text{c}}}_{i,j=1}\bar{V}_{ijij}(\rho^{\text{HF}}),
	\qquad
	\varepsilon^{\text{eff}}_{ii'}
	= T^{\text{re}}_{ii'}
	+ \sum^{N_{\text{c}}}_{j=1}\bar{V}_{iji'j}(\rho^{\text{HF}}),
\end{align}
where the $N_{\text{c}}$ represents the number of nucleons in the core. It should be noted that the given Lagrangian (\ref{Lagrangian}) is obtained by fitting the binding energies of several doubly magic nuclei at the Hartree-Fock level, which means higher corrections to the monopole part of energy have already been absorbed into the phenomenological Lagrangian. Consequently, the higher-order corrections to the core energy, such as the contributions of ladder diagrams, are neglected to avoid double-counting. Moreover, the $\widehat{S}$-box corrections to single-particle energies are also not considered to keep the description of ground state properties of closed-shell nuclei almost unchanged. For the two-body part, we need to calculate the ``bare'' TBMEs and the second-order core-polarization corrections, as well as the folded terms. Similarly, the corrections to the monopole part of TBMEs are removed.

In the following, we focus on the calculation details of effective interactions (\ref{Veff}). Imposing spherical symmetry, the single-particle states $\psi$ are specified by a set of quantum numbers $\alpha \equiv (a, m_{a}) \equiv (\tau_{a}, n_{a}, l_{a}, j_{a}, m_{a})$. In the coupled representation, the particle-particle matrix elements can be obtained as follows:
\begin{align}
	V_{abcd;J}
	= \hat{J}^{-2}\sum_{m_{a}m_{b}m_{c}m_{d}M}
	C^{JM}_{j_{a}m_{a}j_{b}m_{b}}
	C^{JM}_{j_{c}m_{c}j_{d}m_{d}}
	\bar{V}_{\alpha\beta\gamma\delta},
\end{align}
where $\bar{V}_{\alpha\beta\gamma\delta}$ is the uncoupled TBMEs, and $\hat{J}\equiv\sqrt{2J+1}$. Noted that the notation used in this paper is standard \cite{Angular1988}, including the Clebsch-Gordan coefficients and $6j$ symbols.

Furthermore, the $V^{\text{3p1h}}$ correction in the $\widehat{Q}$-box consists of the following four terms:
\begin{align}\label{V3p1h}
	V^{\text{3p1h}}_{abcd;J}
	&= f_{adph}
	\sum_{ph\lambda}
	\lambda^{2}
	\begin{Bmatrix}
		j_{a}&j_{c}&\lambda\\
		j_{d}&j_{b}&J
	\end{Bmatrix}
	\frac{\bar{V}_{hbpd;\lambda}
		\bar{V}_{apch;\lambda}}
	{E_{0}-(\varepsilon_{a}+\varepsilon_{d}-\varepsilon_{h}+\varepsilon_{p})}
	\nonumber\\
	&+ f_{bcph}
	\sum_{ph\lambda}
	\lambda^{2}
	\begin{Bmatrix}
		j_{b}&j_{d}&\lambda\\
		j_{c}&j_{a}&J
	\end{Bmatrix}
	\frac{\bar{V}_{hapc;\lambda}
		\bar{V}_{bpdh;\lambda}}
	{E_{0}-(\varepsilon_{b}+\varepsilon_{c} -\varepsilon_{h}+\varepsilon_{p})}
	\nonumber\\
	&+ f_{acph}
	\sum_{ph\lambda}
	\lambda^{2}
	\begin{Bmatrix}
		j_{a}&j_{d}&\lambda\\
		j_{c}&j_{b}&J
	\end{Bmatrix}
	\frac{\bar{V}_{hbpc;\lambda}
		\bar{V}_{apdh;\lambda}}
	{E_{0}-(\varepsilon_{a}+\varepsilon_{c}-\varepsilon_{h}+\varepsilon_{p})}
	\nonumber\\
	&+ f_{bdph}
	\sum_{ph\lambda}
	\lambda^{2}
	\begin{Bmatrix}
		j_{b}&j_{c}&\lambda\\
		j_{d}&j_{a}&J
	\end{Bmatrix}
	\frac{\bar{V}_{hapd;\lambda}
		\bar{V}_{bpch;\lambda}}
	{E_{0}-(\varepsilon_{b}+\varepsilon_{d} -\varepsilon_{h}+\varepsilon_{p})},
\end{align}
where $E_{0}$ represents the starting energy in $\widehat{Q}$-box, and the factors $f_{3p1h}$, corresponding to different intermediate-states, are expressed as:
\begin{align}
	f_{adph} &= (-1)^{J+j_{c}+j_{d}},
	\qquad
	f_{bcph} = (-1)^{J+j_{a}+j_{b}},
	\nonumber\\
	f_{acph} &= -1,
	\qquad\qquad\qquad
	f_{bdph} = -(-1)^{j_{a}+j_{b}+j_{c}+j_{d}}.
\end{align}
Moreover, the particle-hole matrix elements in Eq. (\ref{V3p1h}) can be obtained as follows:
\begin{align}
	V_{abcd;\lambda} 
	= \hat{\lambda}^{-2}\sum_{m_{a}m_{b}m_{c}m_{d}M}
	(-1)^{j_{b}+m_{b}+j_{c}+m_{c}}
	C^{\lambda M}_{j_{a}m_{a}j_{c}-m_{c}}
	C^{\lambda M}_{j_{d}m_{d}j_{b}-m_{b}}
	\bar{V}_{\alpha\beta\gamma\delta}.
\end{align}
Additionally, the different energy denominators in Eq. (\ref{V3p1h}) can both be simplified as $\varepsilon_{h} - \varepsilon_{p}$, when the model space is degenerate, namely $E_{0}=2\varepsilon_{a}=2\varepsilon_{b}=2\varepsilon_{c}=2\varepsilon_{d}$. 

The $V^{\text{4p2h}}$ correction in the $\widehat{Q}$-box can be evaluated in the coupled representation:
\begin{align}
	V^{\text{4p2h}}_{abcd;J}
	= \sum_{hh'} \frac{\bar{V}_{abhh';J}\bar{V}_{hh'cd;J}}{E_{0}-(\varepsilon_{a}+\varepsilon_{b}+\varepsilon_{c}+\varepsilon_{d}-\varepsilon_{h}-\varepsilon_{h'})},
\end{align}
Similarly, in a degenerate model space, the energy denominator can be further simplified as $\varepsilon_{h}+\varepsilon_{h'} - \frac{1}{2}(\varepsilon_{a}+\varepsilon_{b}+\varepsilon_{c}+\varepsilon_{d})$. Moreover, it should be pointed out that, in this work, we removed the contributions from the isoscalar two-particle cross-shell excitations, which could lead to divergent binding energies with the given Lagrangian (\ref{Lagrangian}). Additionally, the center-of-mass correction is evaluated with the projection-before-variation method \cite{Schmid1991NPA530}.

\vspace{10mm}%
%

\begin{multicols}{2}

\end{multicols}
\end{document}